\documentclass[letter]{aa} 

\usepackage{graphicx}
\usepackage{txfonts}
\usepackage{varwidth} 
\usepackage{color}

\begin{document}

\title{The young exoplanetary system TOI-4562: Confirming the presence
  of a third body in the system}


 \author{
V.~Fermiano \inst{1}
\and
R.~K.~Saito \inst{1}
\and
V.~D.~Ivanov \inst{2}
\and 
C.~Caceres \inst{3}
\and
L.~A.~Almeida \inst{4}
\and
J.~Aires\inst{5}
\and
J.~C.~Beamin \inst{6}
\and
D.~Minniti\inst{3,7}
\and
T.~Ferreira\inst{8}
\and
L.~Andrade\inst{9}
\and
B.~W.~Borges\inst{10}
\and
L.~de~Almeida\inst{9}
\and
F.~Jablonski\inst{11}
\and
W. Schlindwein\inst{11}
}
 \institute{
Departamento  de F\'isica,  Universidade  Federal  de Santa  Catarina,
Trindade 88040-900, Florian\'opolis, Brazil
\and
European  Southern  Observatory,  Karl  Schwarzschildstr  2,  D-85748,
Garching bei München, Germany
\and
Instituto de  Astrof\'isica, Dep.  de Ciencias F\'isicas,  Facultad de
Ciencias  Exactas, Universidad  Andres Bello,  Av. Fern\'andez  Concha
700, Santiago, Chile
\and
Escola de Ci\^encias  e Tecnologia, Universidade Federal  do Rio Grande
do Norte, Campus Universit\'ario, Natal, RN, 59072-970, Brazil
\and
Departamento de F\'isica, Universidade Federal do Rio Grande do Norte,
59072-970, Natal-RN, Brazil
\and
Fundaci\'on Chilena de Astronom\'ia, El Vergel 2252, Santiago, Chile 
\and
Vatican Observatory, Specola Vaticana, V-00120, Vatican City,
Vatican City State
\and
Department of Astronomy, Yale University, 219 Prospect Street,
New Haven, CT 06511, USA
\and
Laborat\'orio Nacional de Astrof\'isica, Rua Estados Unidos 154,
37504-364, Itajub\'a-MG, Brazil
\and
Coordenadoria   Especial  de   F\'isica,  Qu\'imica   e  Matem\'atica,
Universidade Federal de Santa Catarina, Jardim das Avenidas 88906-072,
Ararangu\'a, Brazil
\and
Instituto Nacional  de Pesquisas  Espaciais, Avenida  dos Astronautas,
1758, S\~ao Jos\'e dos Campos-SP, Brazil
}

\offprints{\small{\tt{vitorfreitasfermiano@outlook.com}}}
 
\date{Received September 15, 1996; accepted March 16, 2999}

 
 \abstract
 {Young planetary systems represent  an opportunity to investigate the
   early stages of (exo)planetary  formation because the gravitational
   interactions  have  not  yet   significantly  changed  the  initial
   configuration of the system.}
 {TOI-4562\,b  is  a  highly   eccentric  temperate  Jupiter  analogue
   orbiting a young F7V-type star of $<700$~Myr in age with an orbital
   period of $P_{orb} \sim 225$~days  and an eccentricity of $e=0.76$,
   and is one of the largest known exoplanets to have formed in situ.}
 {We  observed a  new transit  of  TOI-4562\,b using  the 0.6-m  Zeiss
   telescope  at the  Pico  dos Dias  Observatory  (OPD/LNA) in  Minas
   Gerais,  Brazil, and  combine  our data  with Transiting  Exoplanet
   Survey Satellite  (TESS) and  archive data, with  the aim  being to
   improve the ephemerides of this interesting system.}
 {The  $O-C$ diagram  for the  new  ephemeris is  consistent with  the
   presence   of   a  giant   planet   in   an  outer   orbit   around
   TOI-4562. TOI-4562\,c is a planet  with a mass of $M=5.77~M_{Jup}$,
   an orbital period of $P_{orb}= 3990$~days, and a semi-major axis of
   $a = 5.219$~AU.}
 {We  report the  discovery  of TOI-4562\,c,  the  exoplanet with  the
   longest orbital  period discovered to  date via the  transit timing
   variation (TTV) method.   The TOI-4562 system is in  the process of
   violent evolution with intense dynamical changes --- judging by its
   young age and high eccentricity --- and is therefore a prime target
   for studies of formation and evolution of planetary systems.}
 
\keywords{  Planets   and  satellites:   formation  ---   Planets  and
  satellites: dynamical evolution and stability --- Stars: activity }

\maketitle
%

\section{Introduction}

Young stellar systems provide a unique window onto the early stages of
planetary evolution, which  are marked by dynamic  processes and rapid
changes.  These  systems undergo  significant transformations  as they
interact  and  evolve,  and  studying young  systems  offers  valuable
insights into  the formation and  evolution of planets.   By examining
young  stellar systems,  we gain  a fundamental  understanding of  the
mechanisms that  drive planetary formation and  the initial conditions
that shape the diversity of planetary architectures observed in mature
systems.

TOI-4562\,b (=TIC  349576261) is a highly  eccentric temperate Jupiter
analogue           discovered          by           \citet[][hereafter
  HEI23]{2023AJ....165..121H}. The planet orbits a young F7V-type star
of    $<$700   Myr    in    age   with    an    orbital   period    of
$P_{orb}=225.11781^{+0.00025}_{-0.00022}$~days and  an eccentricity of
$e=0.76  \pm 0.02$  (HEI23).  Combining  transits from  the Transiting
Exoplanet  Survey Satellite  (TESS) satellite  with a  transit from  a
ground-based  telescope, and  spectroscopic data,  HEI23 measured  the
planet's radius and mass as $R_P =1.118^{+0.013}_{-0.014}~R_{Jup}$ and
$M_P   =2.30^{+0.48}_{-0.47}~M_{Jup}$.    The  HEI23   ephemeris   for
TOI-4562\,b results in an observed-minus-calculated variation (`$O-C$'
or  `TTV', short  for  transit timing  variation)  at the  $\sim$3--15
minute level for all five  transits used in their analysis, suggesting
the presence of a third body.

We observed a transit of TOI-4562\,b  at the Pico dos Dias Observatory
(OPD/LNA) in Brazil  on November 9 2023, which, much  to our surprise,
occurred $\sim 2$~hours  earlier than expected according  to the HEI23
ephemeris ($O-C=116$~min.).  By combining this new event time with the
existing data from TESS and ground-based observations, we were able to
provide  a  new  ephemeris  for   TOI-4562\,b  whose  TTV  diagram  is
consistent with the  presence of a giant gas planet  in an outer orbit
around  TOI-4562.  Here  we  report the  discovery  of TOI-4562\,c,  a
Jupiter-size planet of $M_P =5.77 M_{Jup}$ with an orbit of $P_{orb} =
3990$~days, and  semi-major axis a $\sim  5.22$ AU. TOI-4562 c  is the
longest-period TTV-discovered exoplanet  to date\footnote{According to
  the  NASA Exoplanet  Archive  on the  time  of writing  ({\small{\tt
      https://exoplanetarchive.ipac.caltech.edu/}})}.

\begin{figure}
 \centering
 \includegraphics[scale=0.53]{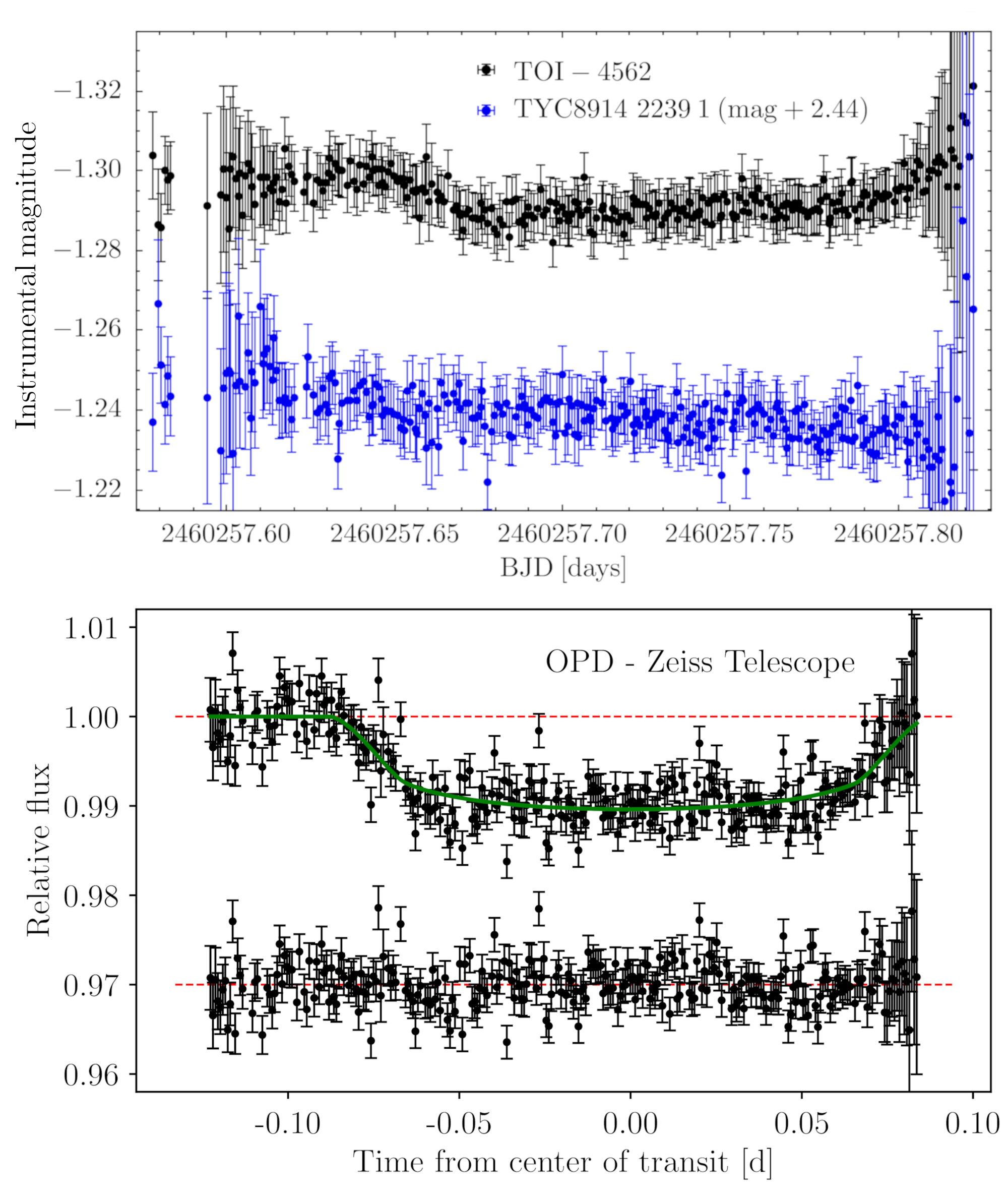}
 \caption {Transit of TOI-4562\,b observed at the OPD/LNA on the night
   of  UT 2023  Nov 9.   Top panel:  Raw data  of TOI-4562\,b  and the
   standard  star  TYC~8914-2239-1  collected  at  the  OPD/LNA.   The
   $y$-axis   is  the   instrumental  magnitude   with  the   data  of
   TYC~8914-2239-1   displaced   by    +2.44~mag   for   visualisation
   purposes. Bottom  panel: Normalised  transit of  TOI-4562\,b (black
   data-points) and  the modelled  transit signal (green  solid line).
   Midtransit occurred at $T_{mid} (BJD) = 2,460,257.73350$, resulting
   in an $O-C$  of $\sim 2$~hours in relation to  the HEI23 ephemeris.
   The  residuals after  the fitting  of  the transit  model are  also
   shown. Fig.~\ref{fig:all-transits} in  the Appendix complements the
   fit  for the  OPD  transit,  showing the  fits  for  the five  TESS
   transits.}
\label{fig:opd}
\end{figure}

\begin{table*}
\caption{Midtransit instances for TOI-4562\,b.}
\begin{center}
\label{tab:obs}
\begin{tabular}{cccrccr}
\hline \hline 
\noalign{\smallskip}
Cycle & Observed & Observatory & ($O-C$)$_{\rm HEI23}$ & ~~~~~~ & Calculated & TTV~~~\, \\
 & BJD [days] & & [min]~~~~~ & &BJD [days] & [min]~~~ \\
\noalign{\smallskip}
\hline
\noalign{\smallskip}
\vspace{1 mm}
0 & $2,458,456.87916^{+0.00112}_{-0.00099}$ & TESS & +10.7712 & & $2,458,456.900401$ & $-$30.7440 \\
\vspace{1 mm}
1 & $2,458,681.98020^{+0.00092}_{-0.00088}$ & TESS & $-$13.3776 & & $2,458,682.006969$ & $-$38.4624 \\
\vspace{1 mm}
3 & $2,459,132.21623^{+0.00110}_{-0.00111}$ & TESS & $-$12.7872 & & $2,459,132.220105$ & $-$5.0112 \\
\vspace{1 mm}
4 & $2,459,357.34510^{+0.00096}_{-0.00095}$ & TESS & +3.1392 & & $2,459,357.326673$ & +27.3456 \\
\vspace{1 mm}
5 & $2,459,582.46321^{+0.00142}_{-0.00142}$ & \,\,LCOGT* & +3.5712 & & $2,459,582.433241$ & +44.2080 \\
\vspace{1 mm}
7 & $2,460,032.64015^{+0.00113}_{-0.00111}$ & TESS & $-$80.9280 & & $2,460,032.646377$ & $-$7.4304 \\
\vspace{1 mm}
8 & $2,460,257.73350^{+0.00369}_{-0.00385}$ & OPD/LNA & $-$116.1504 & & $2,460,257.752945$ & $-$26.2224 \\
\noalign{\smallskip}
\hline
\end{tabular}
\tablefoot{The      cycles     are      numbered     according      to
  $T_0(BJD)=2458456$~days.   The  second  column  shows  the  observed
  midtransits  for  TOI-4562\,b  for  the seven  cycles  used  in  our
  analysis. For the LCOGT observations,  the midtransit is from HEI23.
  The remaining  midtransits are  calculated using  a multiple-transit
  simultaneous-fit procedure  (see Section  4).  Column  four presents
  the $O-C$ according  to the HEI23 ephemeris, while  columns five and
  six present the  calculated midtransit and the  respective TTVs with
  respect to the ephemeris obtained in this work.}
\end{center}
\end{table*}

\section{Ground-based observations of TOI-4562}

Ground-based observations of TOI-4562 were  carried out with the 0.6-m
f/12.5 Zeiss telescope at the Pico dos Dias Observatory on UT November
9 2023.   OPD/LNA is located  in the  Brazilian state of  Minas Gerais
($22^{\rm o}\,32'\,04''$~S, $45^{\rm  o}\,34'\,57''$~W) at an altitude
of  1,864\,m above  sea level.   TOI-4562  is at  coordinates RA,  Dec
(J2000)  =07:28:02.41,   $-$63:31:04,  thus  allowing   good  coverage
starting  at  2023  Nov  9  UT  01:51:33,  with  an  airmass  of  2.45
(corresponding  to  an altitude  of  $23^\circ$),  reaching a  minimum
airmass of 1.32 at UT  $\sim$07:20. The observations were concluded at
2023 Nov 9 UT 07:46:10, with an airmass of 1.35 (altitude $48^\circ$).
For TOI-4562\,b observations, the telescope was equipped with an Andor
iXon  EMCCD camera,  with a  field of  view (FoV)  of approximately  6
$\times$ 6 arcmin.

The observations  of TOI-4562\,b  were conducted  with the  $R$ filter
($\lambda_{\rm C} \sim  6450\,\,\AA$), an exposure time  of 60 seconds
and readout  time of  2.3 seconds,  thus allowing  a cadence  of $\sim
1$~minute (62.3 sec).   This coverage should allow  observation of the
first half of  the TOI-4562\,b transit (cycle \#8,  according to HEI23
ephemeris), which would have its  midtransit near the twilight time in
the  OPD/LNA.   The observations  were  taken  under good  atmospheric
conditions and the  photometric data were reduced  using standard IRAF
tasks to produce the TOI-4562\,b  light curve.  Eight field stars were
used   for   the   differential   photometry,   with   TYC~8914-2239-1
\citep[$V=11.43$~mag and $r=11.08$~mag;][]{2014AJ....148...81M} as the
main reference.  The unreduced data of TOI-4562\,b and TYC~8914-2239-1
collected  at  the   OPD/LNA  are  presented  in  the   top  panel  of
Fig.~\ref{fig:opd},  while the  normalised transit  of TOI-4562\,b  is
shown in  its bottom panel. The  data for the latter  are available in
Table A.1 in the Appendix.

Our updated ephemeris  (discussed in Section 4)  suggests that another
transit of TOI-4562\,b --- cycle  \#9 according to the HEI23 ephemeris
--- was expected  to occur  on 21  June 2024,  with the  midtransit at
08:06:51 UT.  Although  the transit the OPD/LNA was  able to partially
observe the transit, the extremely  low elevation of $\lesssim 10$~deg
prevented data collection.

\begin{figure*}
\centering
\includegraphics[scale=0.75]{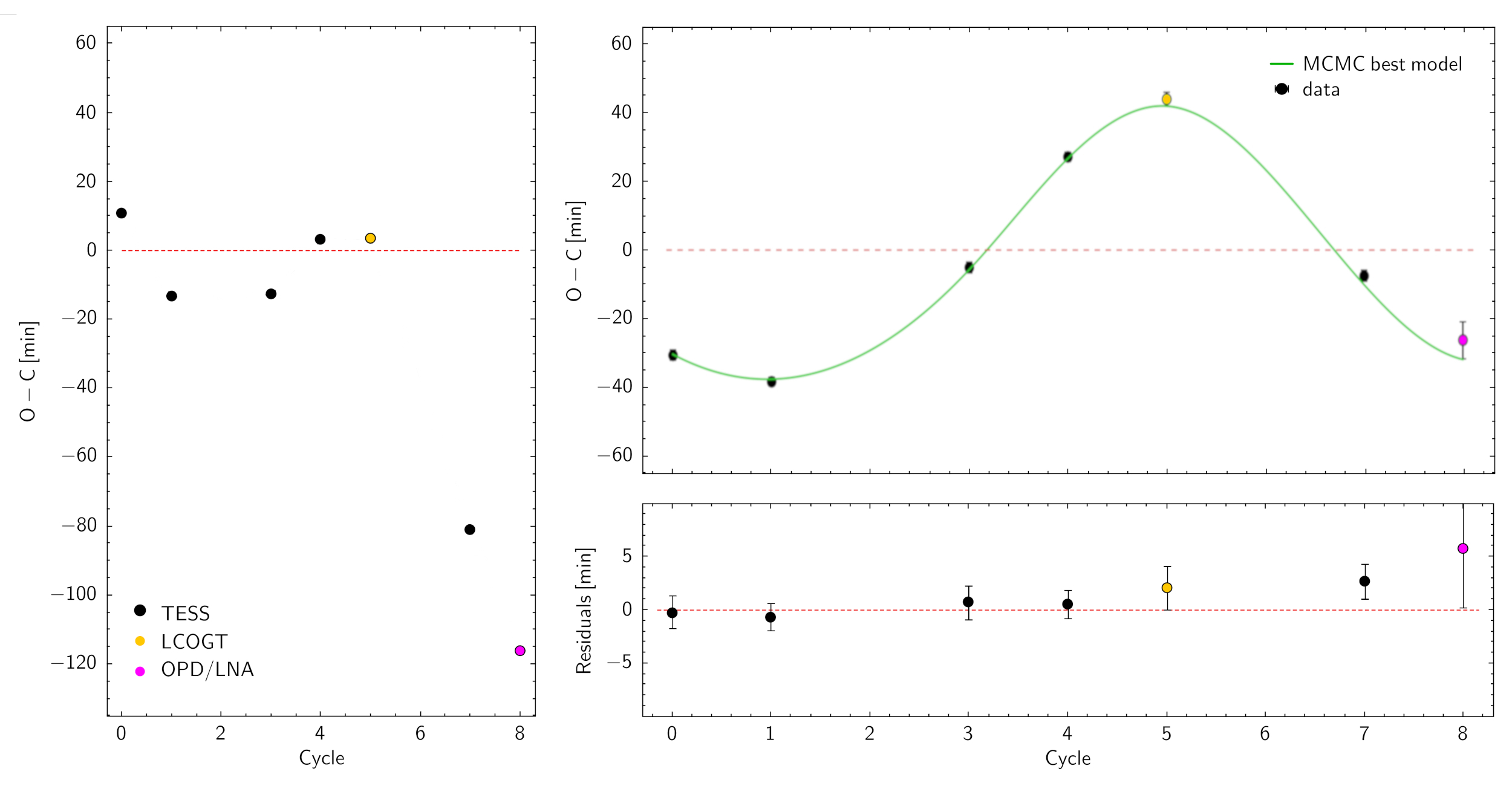}
\caption{$O-C$ diagrams of the  TOI-4562\,b transit. Left panel: $O-C$
  diagram for all  seven events according to the  HEI23 ephemeris.  In
  addition  to the  large  $O-C$  value for  cycle  \#8 observed  from
  OPD/LNA, the $O-C$  for cycle \#7 observed by TESS  is also notable.
  Top  right  panel:  $O-C$   diagram  of  the  TOI-4562\,b  transits,
  according to  the ephemeris  described in  Equation 1.  The observed
  midtransit times  are shown  as filled circles,  while the  model is
  shown  as a  green curve.  Bottom  right panel:  Residuals from  the
  fitting in the top right panel, with the respective error bars.  The
  colour coding of the data points  denotes the observatory and is the
  same in  all panels: black for  TESS, yellow for LCOGT,  and magenta
  for OPD/LNA.}
\label{fig:o-c}
\end{figure*}

\section{Data archive searches}

TOI-4562 lies  in the  TESS southern continuous  viewing zone  and was
photometrically monitored with almost  no interruptions during years 1
and 3  of operations.  The star was  observed at a  cadence of  30 min
during Sectors 1-8 (from July 25 2018 to February 28 2019) and Sectors
10-13 (March 26  2019 to July 18  2019), at a cadence of  2 min during
Sectors 27–39 (from July 4 2020  to June 24 2021), and at a
cadence of 20\,sec in  sector 63 (March 10 2023 to  April 6 2023). The
first  transit  signature of  TOI-4562\,b  was  detected by  the  TESS
Science Processing Operations  Center \citep{2016SPIE.9913E..3EJ}, and
the object was designated as a TESS Object of Interest (TOI) in 2021.

We   retrieved   all   available   TESS  data   for   TOI-4562   using
\texttt{lightkurve} \citep{2018ascl.soft12013L},  which were processed
with      the      SPOC      pipeline      \citep{2016SPIE.9913E..3EJ,
  2020DPS....5220703J}.  A total of  five transits of TOI-4562\,b were
found in  TESS light curves,  including a  new event that  occurred on
March 29  2023 (cycle  \#7).  The  four previous  events were  used by
HEI23 in their analysis (cycles \#0, \#1, \#3 and \#4).  The timing of
a TOI-4562\,b transit observed on January  3 2022 at the LCOGT's South
African Astronomical Observatory (SAAO) node  using two 1 m telescopes
(details in HEI23) was also included in our analysis.

\begin{figure}
\centering
 \includegraphics[scale=1.2]{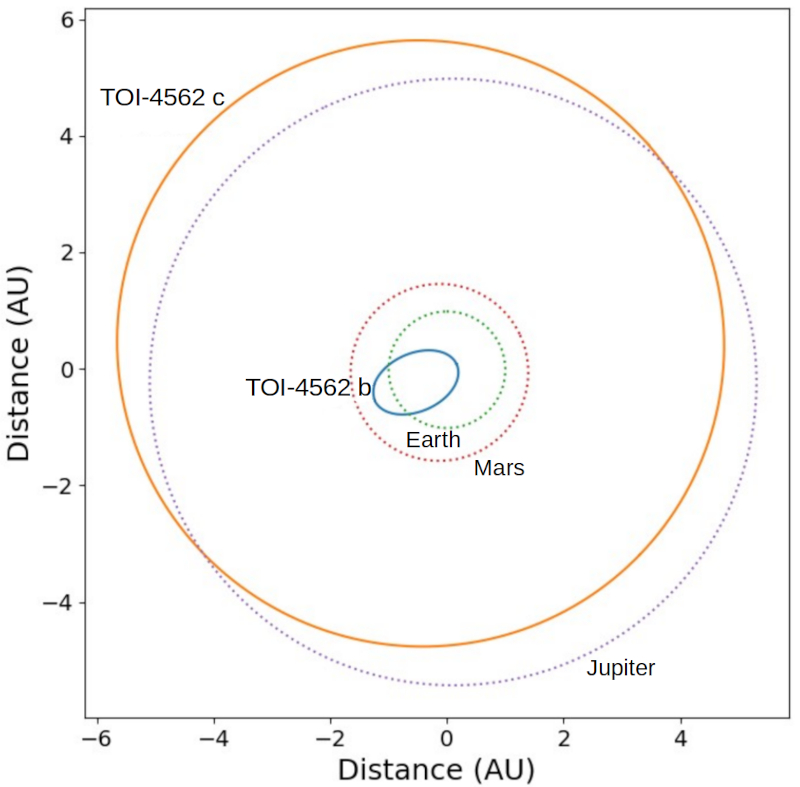}
 \caption{Schematic  diagram showing  the  orbits  of TOI-4562\,b  and
   TOI-4562\,c compared with planets  of the Solar System. TOI-4562\,c
   is  a  giant  planet  with  5.73 Jupiter  masses  in  an  orbit  with a
   semi-major axis of $5.22$~AU and an orbital period of 3990~days.}
 \label{fig:orbits}
\end{figure}

\section{Data analysis}

The aim  of our analysis is  to understand the variation  of the $O-C$
detected  in  the  TOI-4562\,b  transits,  which  could  indicate  the
presence  of  a third  body  in  the  system.   We first  performed  a
transiting exoplanet  light-curve fitting  on the  Zeiss and  the five
TESS  light curves  simultaneously.   We modelled  the transit  signal
using the \texttt{batman} package \citep{2015PASP..127.1161K} together
with a Markov Chain Monte Carlo (MCMC) method, implemented through the
\texttt{emcee} package \citet{2013PASP..125..306F}.   We considered as
free parameters  the planet-to-star  radius $R_{p,b}/R_\star$  for the
Zeiss  and   TESS  light  curves  $k_V$   and  $k_\mathrm{TESS}$,  the
semi-major axis normalised to  the stellar radius $a_{b}/R_\star$, the
inclination angle $i$,  and the transit times  $t_\mathrm{c}$ for each
epoch individually.  We obtained all  the other parameters (with their
uncertainties) from HEI23,  which were kept fixed  during the fitting.
This  includes two  quadratic  limb-darkening  coefficients $u_1$  and
$u_2$   in   the   V-filter    that   were   interpolated   from   the
\citet{2000A&A...363.1081C} tables.   In addition, we included  in the
fit a second-order polynomial to  account for the normalisation of the
light curves. For  the MCMC fit, we used 50,000  steps and 64 walkers,
and  we  assumed  uniform  priors  for  all  the  free  parameters.The
parameters  are  summarised in  Tables  1  and  2 (transit  times  and
remaining  parameters, respectively),  and  show  good agreement  with
those presented in  HEI23. We find that all the  parameters in the fit
are in agreement with those presented in HEI23.

Our  next step  consisted of  determining  the orbital  period of  the
system  and performing  a  polynomial  fit to  obtain  the delays  and
advances in  the orbital cycles  of the  exoplanet and thus  build the
$O-C$ diagram.  Using the midtransit  timing for all seven events (see
Table 1), a new ephemeris for TOI-4562\,b was obtained, as follows:

\begin{equation}
T_{mid} (BJD)= 2459476.90051^{+0.00053}_{-0.00057} + 225.10640^{+0.00015}_{-0.00015} \times E. 
\end{equation}

A  longer  baseline produced  a  slightly  different ephemeris  with respect
to  that obtained by HEI23.  In particular, we  note that the orbital period is
approximately 16~min ($=0.00114$~days) shorter than the previous HEI23
value,  thus  producing   a  distinct  $O-C$  diagram,   as  shown  in
Fig.~\ref{fig:o-c}. When  fitting the transit-time measurements  using a
linear  ephemeris,  we  note  a possible  periodic  signal  with  a
semi-amplitude of approximately 40 min. This signal could indicate the
presence of an additional  third body gravitationally interacting with
the system.

\begin{table}
\caption{Parameters of TOI-4562\,b and TOI-4562\,c}
\begin{center}
\label{tab:obs}
\begin{tabular}{l@{ }p{3,5cm}r@{ }}
\hline \hline 
\noalign{\smallskip}
Parameter & Description & Value\\
\noalign{\smallskip}
\hline
\noalign{\smallskip}
\vspace{1 mm}
\textbf{TOI-4562\,b}   &                                    &                         \\
\vspace{1 mm}
$k_\mathrm{OPD/LNA}$    & Planet-to-star radius                   & $0.0960^{+0.0071}_{-0.0075}$  \\
\vspace{1 mm}
$k_\mathrm{TESS}$       & Planet-to-star radius                   & $0.09485^{+0.0015}_{-0.0015}$ \\
\vspace{1 mm}
$a/R_\mathrm{s}$        & Normalised semi-major axis       & $150.3^{+12.4}_{\,\,\,-9.8}$  \\
\vspace{1 mm}
$i$                   & Orbit inclination [deg] & $+89.15^{+0.21}_{-0.17}$ \\
\vspace{1 mm}
$u_1$                 & Limb-darkening coefficient & 0.41 \\
\vspace{1 mm}
$u_2$                 & Limb-darkening coefficient & 0.29 \\
\vspace{1 mm}
$T_0$                 & Linear ephemeris [BJD]             & $2459476.90040^{+0.00051}_{-0.00059}$  \\
\vspace{1 mm}
$P_{orb}$              & Orbital period [days]              & $225.10640^{+0.00015}_{-0.00015}$      \\
\vspace{1 mm}
$\Omega_{b}$          & Longitude of the ascending  node [deg] & $24^{+17}_{-16}$     \\
\vspace{1 mm}
$\mathcal{M}_{b}$     &  Mean anomaly [deg]                 & $ 38.297 \pm 0.004 $     \\
                     &                                      &                          \\
\vspace{1 mm}
\textbf{TOI-4562\,c}  &                                      &                          \\
\vspace{1 mm}
$M_P$                & Mass Planet [M$_{jup}$]                & $5.77^{+0.37}_{-0.56}$       \\
\vspace{1 mm}
$P_{orb}$             & Orbital period [days]                & $3990^{+201}_{-192}$  \\
\vspace{1 mm}
$e$                 & Eccentricity                          & $0.122^{+0.027}_{-0.026}$      \\
\vspace{1 mm}
$a$                 & Semimajor axis [AU]                   & $5.219 \pm 0.002$        \\
\vspace{1 mm}
$\omega$            & Argument of periapsis [deg]  & $3^{+20}_{-21}$      \\
\vspace{1 mm}
$i$                 & Orbit inclination [deg]      & $90.00^{+0.17}_{-0.19}$    \\
\vspace{1 mm}
$\Omega_{c}$         & Longitude of the ascending node [deg] & $-57^{+23}_{-24}$    \\
\vspace{1 mm}
$\mathcal{M}_{c}$   & Mean anomaly [deg]                     & $194^{+24}_{-23}$    \\
\noalign{\smallskip}
\hline
\end{tabular}
\tablefoot{Orbital parameters  of TOI-4562\,b and  TOI-4562\,c derived
  from our analysis.  A corner plot for the  posterior distribution of
  our  free parameters  is presented  in Fig.~\ref{fig:corner}  of the
  Appendix.}
\end{center}
\end{table}

We  searched for  the  parameters  of this  possible  third object  by
modelling  the observed  transit-times with  the \texttt{TTVFast}  code
\citep[]{2014ApJ...787..132D}, which  is a modified  n-body simulation
code  that uses  a  WH integrator  \citep{1991AJ....102.1528W} with  a
Keplerian orbit  interpolator to calculate  transit times and  TTVs in
transiting                      planetary                      systems
\citep[]{2014ApJ...787..132D,2021MNRAS.507.1582A}.  The code  requires the  parameters of
both  bodies,  the  mass  of the  star,  and the  time and  the
time-step  for  the  integration.   We  took  the  TOI-4562  mass  and
TOI-4562\,b  parameters  given  by   HEI23,  and  integrated  from  BJD
$2458254$  to  BJD $2460258$  with  a  time-step  of 0.01  days.   The
parameters  of the  third body and  those of  TOI-4562\,b that  do not
appear in HEI23  were taken as free parameters and  were determined using a
MCMC procedure  from the \texttt{emcee}  package.  At each step  of the
MCMC, we used the generated values to compute the transit times of the
observed planet  and compared them  with the observed times  through a
$\chi^{2}$ goodness-of-fit estimation, that is,

\begin{equation}
\chi^{2} = \sum_{i}^{n} \left( \frac{O_{i} - C_{i}}{\sigma_{i}} \right)^{2},
\end{equation}
where $O_{i}$  are the observed  times, $C_{i}$ are  the corresponding
computed times, and $\sigma_{i}$ are the errors in the observations.

As free parameters,  we use the mass of the  third body ($M_{c}$), the
orbital period ($P_{c}$), the eccentricity ($e_{c}$), the longitude of
the   ascending   node   ($\Omega_{c}$),   the   periastron   argument
($\omega_{c}$), and the mean anomaly ($\mathcal{M}{c}$), together with
the  longitude  of the  ascending  node  ($\Omega_{b}$) and  the  mean
anomaly  ($\mathcal{M}_{b}$)  of  the  transiting  body.  We  manually
searched for the  best initial values of the free  parameters to start
the MCMC.   The priors  were distributed  normally around  the initial
values. We used  16 walkers (twice the number of  free parameters) and
5,000 steps, discarding the first  100 steps as burn-in. Table~2 lists
the median values  and the 1$\sigma$ confidence  interval ($\pm 34\%$)
obtained from the a posterior  distributions. The corner plot with the
posterior   distributions    of   the    parameters   is    shown   in
Fig.~\ref{fig:corner} in the Appendix.

Figure~\ref{fig:o-c}  shows  the  best solution  for  the  midtransits
together  with the  observed  transit times.  The  next five  upcoming
events   of   TOI-4562\,b  are   presented   in   the  Appendix   (see
Table~\ref{tab:obs}                                                and
Fig.~\ref{fig:predict}).  Figure~\ref{fig:orbits} shows  the schematic
diagram  of the  TOI-4562 system  compared with  planets of  the Solar
System.

\section{Discussion}

Young planetary systems  such as TOI-4562 represent  an opportunity to
observe the early stages of  planetary formation. In the first million
years of  the life of  these systems, gravitational  interactions have
not   yet   significantly    changed   their   initial   configuration
\citep[e.g.][]{2021A&A...649A.177M}.  Figure~\ref{fig:orbits}  shows a
schematic diagram of the TOI-4562  system compared with planets of the
Solar System.  TOI-4562\,c is the planet with the largest period found
so far  using the  TTV technique.  The large TTV  amplitude may  be an
indication of  planetary migration, although the  ratio between orbits
is  very high  and there  is no  indication of  resonance between  the
planets $(P_{\rm c}/P_{\rm b} \sim  17.7)$.  In addition, the orbit of
TOI-4562\,b  is  one  of  the most  eccentric  among  exoplanets  with
indications of in situ formation \citep[e.g.][]{2014prpl.conf..619C}.

The TTV method  is susceptible to dynamic  degeneracies between planet
eccentricity   and   planet   mass,  and   perturbations   caused   by
non-transiting  planets can  be  misattributed  to transiting  planets
\citep{2016PhDT.......181W}.   This  degeneracy  can  be  broken  with
additional information such as radial-velocity measurements, as in the
analysis of  TOI-4562\,b by HEI23;  or the assumption of  stability of
the                          planetary                          system
\citep{2012ApJ...750..114F,2012ApJ...761..122L,2017AJ....153..265W},
which  is unlikely  here  given the  youth of  TOI-4562.   In our  TTV
analysis,  we kept  the  mass  of TOI-4562\,b  as  a  prior and  fixed
parameter with the  value (and uncertainties) as reported  by HEI23 to
minimise this  degeneracy effect,  thus decreasing the  probability of
solutions  with  similar  residuals for  different  configurations  of
planet mass and eccentricity for TOI-4562\,c.

Other  possible interpretations  of  the TTV  signal  may include  the
consideration  that TOI-4562  is  made up  of only  the  star and  the
transiting  planet. Firstly,  there  is the  possibility  that TTV  is
caused by  stellar activity. The most  plausible explanation entailing
dynamical interaction  with the star,  causing the TTV signal,  is the
Applegate effect \citep{1992ApJ...385..621A} applied to exoplanets, as
described in  \citet{2010MNRAS.405.2037W}.  According to  this effect,
the star undergoes a magnetic cycle that alters the gravitational pull
of its  rotational bulge on  the planet, causing slight  variations in
its orbital period.  To produce the observed TTV  signal, the magnetic
cycle would need to have a period of $\sim 1995$~days (the initial fit
for the TTV signal).  For magnetic dynamos typical of solar-type stars
like Kepler-19,  \citet{2010MNRAS.405.2037W} calculate  TTV variations
of less than  1 second over timescales of several  years; too small to
explain our data. In addition  to this dynamical interaction, activity
could  cause  an  apparent  TTV  due  to  the  planet's  transit  over
starspots,  and the  rotation of  the  apsidal line  of the  planetary
orbit.

Despite  the high  inclination of  the orbit  of TOI-4562\,c,  we were
unable  to find  any  signal  of its  transit  in  the available  TESS
data.  This absence  is not  surprising  given that  the current  TESS
baseline of $\sim 6$~years is shorter than the planet's orbital period
of  almost  11  years.  Moreover, TESS  does  not  provide  continuous
coverage of the TOI-4562 field. Even for planet TOI-4562\,b, which has
a shorter period, only half of the transits have been observed by TESS
since the first detection. In addition, some of the orbital parameters
that should predict the transits of planet TOI-4562\,c still have high
uncertainties due to the small  number of transits observed for planet
TOI-4562\,b,  from  which we  obtained  the  TTV curve.   This  result
highlights  the challenges  in observing  long-period exoplanets  with
current transit-based survey missions.  Future monitoring is necessary
to refine the orbital parameters for  both planets in order to confirm
the presence of transits for TOI-4562\,c.

\section{Conclusions}

We  observed  a new  transit  of  TOI-4562\,b  using the  0.6-m  Zeiss
telescope at the Pico dos Dias Observatory in Minas Gerais, Brazil, in
order to improve the ephemeris  of this interesting system.  The $O-C$
diagram  considering  our  observation,  together with  all  the  other
transits observed  with TESS  and the LCOGT  transit available  in the
literature, allowed us  to detect the presence of a  giant planet in an
outer orbit around TOI-4562. We report the discovery of TOI-4562\,c,
a planet with  a mass of $M=5.77 M_{Jup}$ with an  orbital period of $P_{orb}=
3990$~days, and  a semi-major axis of $a =  5.219$~AU.  To the best of our knowledge,
this is the  exoplanet with the largest orbital period discovered
using  the  TTV  method  known to date.  There  is  also  no sign  of
resonance  between the    orbits of the planets, with  the TOI-4562\,b  orbit
being one of  the most eccentric known. The TOI-4562  system is in the
process of violent  evolution and intense dynamical  changes --- judging
by its young age and the high eccentricity of the TOI-4562\,b planet ---
and is  therefore a prime  target for  studies of the  formation and
evolution of planetary systems.

\begin{acknowledgements}
This  work uses  observations made  at  the Observatorio  do Pico  dos
Dias/LNA  (Brazil).  Funding for the TESS mission is provided by NASA's Science Mission directorate. V.F. acknowledges  support  from CNPq/Brazil and Universidade Federal de Santa Catarin (UFSC) through programs BIP/UFSC and PIBIC/UFSC. R.K.S.   acknowledges  support  from  CNPq/Brazil
through  project  308298/2022-5. R.K.S., L.A.A. and F.J. acknowledge  
support  from CNPq/Brazil through  project 421034/2023-8.
C.C.   acknowledges   support   by  ANID   BASAL   project   FB210003.
D.M. acknowledges support by Fondecyt Project No.  1220724, and by the
Center  for  Astrophysics  and   Associated  Technologies  ANID  BASAL
projects ACE210002 and FB210003.
\end{acknowledgements}

%
%

\clearpage

\begin{appendix}

\section{Complementary material}

\begin{table*}
\caption{Normalised light-curve of the TOI-4562\,b transit observed at
  the  OPD/LNA on  2023 Nov  9.  Time  corresponds to  BJD$-$2460257.}
\small
\begin{center}
\label{tab:obs}
\begin{tabular}{cc|cc|cc|cc}
\hline \hline 
\noalign{\smallskip}
Time [days] & Relative Flux & Time [days] & Relative Flux & Time [days] & Relative Flux & Time [days] & Relative Flux \\
\noalign{\smallskip}
\hline
\noalign{\smallskip}
0.610618  &  1.0008$\pm$0.0036  &  0.666149  &  0.9997$\pm$0.0019  &  0.720236  &  0.9881$\pm$0.0019  &  0.769274  & 0.9908$\pm$0.0018  \\
0.611340  &  0.9965$\pm$0.0037  &  0.667591  &  0.9919$\pm$0.0019  &  0.720957  &  0.9933$\pm$0.0019  &  0.769995  & 0.9896$\pm$0.0018  \\
0.612061  &  1.0005$\pm$0.0036  &  0.668312  &  0.9926$\pm$0.0019  &  0.721678  &  0.9916$\pm$0.0019  &  0.770716  & 0.9908$\pm$0.0018  \\
0.612782  &  0.9981$\pm$0.0032  &  0.669033  &  0.9911$\pm$0.0019  &  0.722399  &  0.9923$\pm$0.0019  &  0.771437  & 0.9894$\pm$0.0018  \\
0.613503  &  0.9975$\pm$0.0030  &  0.669754  &  0.9903$\pm$0.0019  &  0.723121  &  0.9916$\pm$0.0019  &  0.772159  & 0.9889$\pm$0.0019  \\
0.614224  &  0.9994$\pm$0.0028  &  0.670475  &  0.9869$\pm$0.0019  &  0.723842  &  0.9916$\pm$0.0018  &  0.772880  & 0.9877$\pm$0.0018  \\
0.614945  &  1.0004$\pm$0.0026  &  0.671197  &  0.9909$\pm$0.0019  &  0.724563  &  0.9918$\pm$0.0018  &  0.774322  & 0.9889$\pm$0.0018  \\
0.615666  &  0.9949$\pm$0.0025  &  0.671918  &  0.9918$\pm$0.0019  &  0.725284  &  0.9887$\pm$0.0018  &  0.775043  & 0.9893$\pm$0.0018  \\
0.616388  &  0.9978$\pm$0.0024  &  0.673360  &  0.9899$\pm$0.0019  &  0.726005  &  0.9906$\pm$0.0018  &  0.775764  & 0.9896$\pm$0.0018  \\
0.617109  &  1.0071$\pm$0.0024  &  0.674081  &  0.9899$\pm$0.0019  &  0.726726  &  0.9892$\pm$0.0019  &  0.776485  & 0.9890$\pm$0.0018  \\
0.617830  &  0.9945$\pm$0.0023  &  0.674802  &  0.9894$\pm$0.0019  &  0.727447  &  0.9889$\pm$0.0019  &  0.777206  & 0.9915$\pm$0.0018  \\
0.618551  &  1.0029$\pm$0.0022  &  0.675523  &  0.9932$\pm$0.0019  &  0.728168  &  0.9892$\pm$0.0018  &  0.777928  & 0.9960$\pm$0.0019  \\
0.619273  &  1.0010$\pm$0.0022  &  0.676244  &  0.9898$\pm$0.0019  &  0.728890  &  0.9920$\pm$0.0018  &  0.778649  & 0.9923$\pm$0.0018  \\
0.619994  &  0.9995$\pm$0.0022  &  0.676966  &  0.9916$\pm$0.0018  &  0.729611  &  0.9868$\pm$0.0018  &  0.779370  & 0.9860$\pm$0.0018  \\
0.622878  &  0.9968$\pm$0.0021  &  0.677687  &  0.9884$\pm$0.0019  &  0.730332  &  0.9883$\pm$0.0018  &  0.780091  & 0.9873$\pm$0.0018  \\
0.623599  &  1.0003$\pm$0.0021  &  0.679129  &  0.9920$\pm$0.0019  &  0.731053  &  0.9894$\pm$0.0019  &  0.780812  & 0.9905$\pm$0.0018  \\
0.624320  &  1.0006$\pm$0.0022  &  0.679850  &  0.9872$\pm$0.0019  &  0.731775  &  0.9896$\pm$0.0018  &  0.781533  & 0.9888$\pm$0.0018  \\
0.625763  &  0.9944$\pm$0.0021  &  0.680571  &  0.9859$\pm$0.0019  &  0.732496  &  0.9919$\pm$0.0019  &  0.782254  & 0.9910$\pm$0.0018  \\
0.627205  &  0.9996$\pm$0.0021  &  0.681292  &  0.9894$\pm$0.0019  &  0.733216  &  0.9904$\pm$0.0019  &  0.782975  & 0.9929$\pm$0.0018  \\
0.627926  &  0.9989$\pm$0.0021  &  0.682014  &  0.9880$\pm$0.0019  &  0.733938  &  0.9905$\pm$0.0019  &  0.783696  & 0.9889$\pm$0.0019  \\
0.628647  &  0.9972$\pm$0.0021  &  0.683456  &  0.9932$\pm$0.0019  &  0.734659  &  0.9879$\pm$0.0019  &  0.784417  & 0.9891$\pm$0.0018  \\
0.629368  &  0.9990$\pm$0.0021  &  0.684177  &  0.9853$\pm$0.0019  &  0.735380  &  0.9899$\pm$0.0019  &  0.785138  & 0.9876$\pm$0.0018  \\
0.630089  &  1.0011$\pm$0.0021  &  0.684898  &  0.9931$\pm$0.0019  &  0.736101  &  0.9905$\pm$0.0019  &  0.785860  & 0.9953$\pm$0.0018  \\
0.630811  &  1.0045$\pm$0.0021  &  0.686341  &  0.9940$\pm$0.0019  &  0.736823  &  0.9889$\pm$0.0019  &  0.786581  & 0.9955$\pm$0.0018  \\
0.631532  &  1.0001$\pm$0.0021  &  0.687062  &  0.9885$\pm$0.0019  &  0.737544  &  0.9919$\pm$0.0019  &  0.787302  & 0.9889$\pm$0.0018  \\
0.632253  &  1.0032$\pm$0.0020  &  0.687783  &  0.9888$\pm$0.0019  &  0.738265  &  0.9918$\pm$0.0018  &  0.788023  & 0.9865$\pm$0.0019  \\
0.632974  &  1.0020$\pm$0.0020  &  0.688504  &  0.9877$\pm$0.0019  &  0.738986  &  0.9881$\pm$0.0019  &  0.788744  & 0.9900$\pm$0.0018  \\
0.633696  &  1.0017$\pm$0.0020  &  0.689225  &  0.9902$\pm$0.0019  &  0.739707  &  0.9892$\pm$0.0019  &  0.789466  & 0.9919$\pm$0.0018  \\
0.634417  &  1.0017$\pm$0.0020  &  0.689946  &  0.9920$\pm$0.0019  &  0.740428  &  0.9888$\pm$0.0019  &  0.790187  & 0.9928$\pm$0.0021  \\
0.635859  &  0.9980$\pm$0.0020  &  0.690667  &  0.9898$\pm$0.0019  &  0.741150  &  0.9897$\pm$0.0019  &  0.790907  & 0.9921$\pm$0.0019  \\
0.636580  &  1.0037$\pm$0.0020  &  0.691389  &  0.9893$\pm$0.0019  &  0.741871  &  0.9927$\pm$0.0019  &  0.791629  & 0.9892$\pm$0.0018  \\
0.638023  &  0.9992$\pm$0.0020  &  0.692831  &  0.9913$\pm$0.0019  &  0.742592  &  0.9873$\pm$0.0019  &  0.792350  & 0.9903$\pm$0.0019  \\
0.638744  &  0.9972$\pm$0.0019  &  0.693552  &  0.9959$\pm$0.0019  &  0.743313  &  0.9915$\pm$0.0018  &  0.793071  & 0.9907$\pm$0.0018  \\
0.639465  &  1.0027$\pm$0.0020  &  0.694273  &  0.9900$\pm$0.0019  &  0.744034  &  0.9920$\pm$0.0018  &  0.793792  & 0.9925$\pm$0.0018  \\
0.640907  &  0.9985$\pm$0.0021  &  0.694994  &  0.9912$\pm$0.0019  &  0.744755  &  0.9864$\pm$0.0018  &  0.794513  & 0.9918$\pm$0.0019  \\
0.641628  &  1.0026$\pm$0.0020  &  0.696437  &  0.9929$\pm$0.0019  &  0.745476  &  0.9931$\pm$0.0018  &  0.795235  & 0.9912$\pm$0.0019  \\
0.642350  &  0.9997$\pm$0.0019  &  0.697158  &  0.9838$\pm$0.0019  &  0.746198  &  0.9924$\pm$0.0019  &  0.795956  & 0.9923$\pm$0.0019  \\
0.643071  &  1.0045$\pm$0.0020  &  0.697879  &  0.9926$\pm$0.0019  &  0.746919  &  0.9881$\pm$0.0019  &  0.796677  & 0.9919$\pm$0.0019  \\
0.643792  &  1.0013$\pm$0.0020  &  0.698600  &  0.9906$\pm$0.0019  &  0.747640  &  0.9885$\pm$0.0018  &  0.797398  & 0.9921$\pm$0.0019  \\
0.644513  &  1.0018$\pm$0.0020  &  0.699321  &  0.9886$\pm$0.0019  &  0.748361  &  0.9934$\pm$0.0019  &  0.798841  & 0.9904$\pm$0.0020  \\
0.645234  &  0.9981$\pm$0.0020  &  0.700043  &  0.9916$\pm$0.0019  &  0.749082  &  0.9918$\pm$0.0019  &  0.799562  & 0.9933$\pm$0.0020  \\
0.645955  &  1.0018$\pm$0.0020  &  0.700764  &  0.9899$\pm$0.0019  &  0.750524  &  0.9888$\pm$0.0018  &  0.800283  & 0.9931$\pm$0.0021  \\
0.646677  &  0.9985$\pm$0.0019  &  0.701485  &  0.9885$\pm$0.0019  &  0.751246  &  0.9880$\pm$0.0018  &  0.801004  & 0.9937$\pm$0.0021  \\
0.647398  &  0.9976$\pm$0.0020  &  0.702206  &  0.9883$\pm$0.0019  &  0.751967  &  0.9882$\pm$0.0019  &  0.801725  & 0.9993$\pm$0.0022  \\
0.648119  &  1.0012$\pm$0.0020  &  0.702927  &  0.9901$\pm$0.0019  &  0.752688  &  0.9924$\pm$0.0019  &  0.802446  & 0.9911$\pm$0.0023  \\
0.648840  &  1.0027$\pm$0.0020  &  0.703649  &  0.9928$\pm$0.0019  &  0.753409  &  0.9970$\pm$0.0019  &  0.803167  & 0.9934$\pm$0.0023  \\
0.649561  &  1.0001$\pm$0.0020  &  0.704370  &  0.9926$\pm$0.0019  &  0.754130  &  0.9894$\pm$0.0019  &  0.803889  & 0.9934$\pm$0.0024  \\
0.651004  &  0.9988$\pm$0.0019  &  0.705091  &  0.9910$\pm$0.0019  &  0.754851  &  0.9913$\pm$0.0018  &  0.804610  & 0.9969$\pm$0.0026  \\
0.651725  &  0.9993$\pm$0.0019  &  0.705812  &  0.9895$\pm$0.0019  &  0.755573  &  0.9891$\pm$0.0018  &  0.805331  & 0.9954$\pm$0.0027  \\
0.652446  &  0.9994$\pm$0.0019  &  0.706533  &  0.9984$\pm$0.0019  &  0.756294  &  0.9932$\pm$0.0019  &  0.806052  & 0.9996$\pm$0.0028  \\
0.653167  &  0.9977$\pm$0.0019  &  0.707254  &  0.9907$\pm$0.0019  &  0.757015  &  0.9930$\pm$0.0019  &  0.806773  & 0.9988$\pm$0.0031  \\
0.653889  &  0.9948$\pm$0.0020  &  0.707975  &  0.9928$\pm$0.0019  &  0.757736  &  0.9918$\pm$0.0019  &  0.807494  & 0.9926$\pm$0.0033  \\
0.654610  &  0.9988$\pm$0.0020  &  0.708697  &  0.9859$\pm$0.0019  &  0.758457  &  0.9946$\pm$0.0022  &  0.808215  & 0.9928$\pm$0.0035  \\
0.655331  &  0.9965$\pm$0.0020  &  0.709418  &  0.9852$\pm$0.0019  &  0.759178  &  0.9911$\pm$0.0019  &  0.808937  & 0.9956$\pm$0.0039  \\
0.656052  &  0.9960$\pm$0.0020  &  0.710139  &  0.9888$\pm$0.0019  &  0.759899  &  0.9882$\pm$0.0019  &  0.809658  & 0.9973$\pm$0.0043  \\
0.656774  &  0.9967$\pm$0.0019  &  0.711581  &  0.9907$\pm$0.0019  &  0.761342  &  0.9923$\pm$0.0018  &  0.810379  & 0.9974$\pm$0.0045  \\
0.657494  &  0.9901$\pm$0.0019  &  0.712302  &  0.9917$\pm$0.0019  &  0.762063  &  0.9909$\pm$0.0018  &  0.811100  & 0.9992$\pm$0.0049  \\
0.658216  &  0.9970$\pm$0.0020  &  0.713024  &  0.9890$\pm$0.0019  &  0.762784  &  0.9873$\pm$0.0018  &  0.811822  & 0.9967$\pm$0.0056  \\
0.658937  &  0.9961$\pm$0.0022  &  0.714466  &  0.9886$\pm$0.0019  &  0.763505  &  0.9895$\pm$0.0018  &  0.812543  & 1.0004$\pm$0.0061  \\
0.659658  &  1.0041$\pm$0.0024  &  0.715187  &  0.9909$\pm$0.0018  &  0.764226  &  0.9877$\pm$0.0018  &  0.813264  & 0.9993$\pm$0.0068  \\
0.660379  &  0.9939$\pm$0.0029  &  0.715908  &  0.9902$\pm$0.0018  &  0.764948  &  0.9883$\pm$0.0018  &  0.813985  & 0.9987$\pm$0.0071  \\
0.661100  &  0.9980$\pm$0.0031  &  0.716630  &  0.9878$\pm$0.0019  &  0.765669  &  0.9902$\pm$0.0018  &  0.814707  & 0.9935$\pm$0.0078  \\
0.662543  &  0.9961$\pm$0.0020  &  0.717351  &  0.9890$\pm$0.0019  &  0.766390  &  0.9874$\pm$0.0018  &  0.815428  & 1.0070$\pm$0.0090  \\
0.663264  &  0.9941$\pm$0.0019  &  0.718072  &  0.9850$\pm$0.0019  &  0.767111  &  0.9911$\pm$0.0018  &  0.816149  & 1.0019$\pm$0.0095  \\
0.663985  &  0.9950$\pm$0.0019  &  0.718793  &  0.9889$\pm$0.0019  &  0.767832  &  0.9907$\pm$0.0018  &  0.816870  & 1.0001$\pm$0.0109  \\
0.664706  &  0.9936$\pm$0.0020  &  0.719514  &  0.9894$\pm$0.0019  &  0.768553  &  0.9894$\pm$0.0018  &            &                    \\
\noalign{\smallskip}
\hline
\end{tabular}
\end{center}
\end{table*}

\begin{figure}
\centering
\includegraphics[scale=0.75]{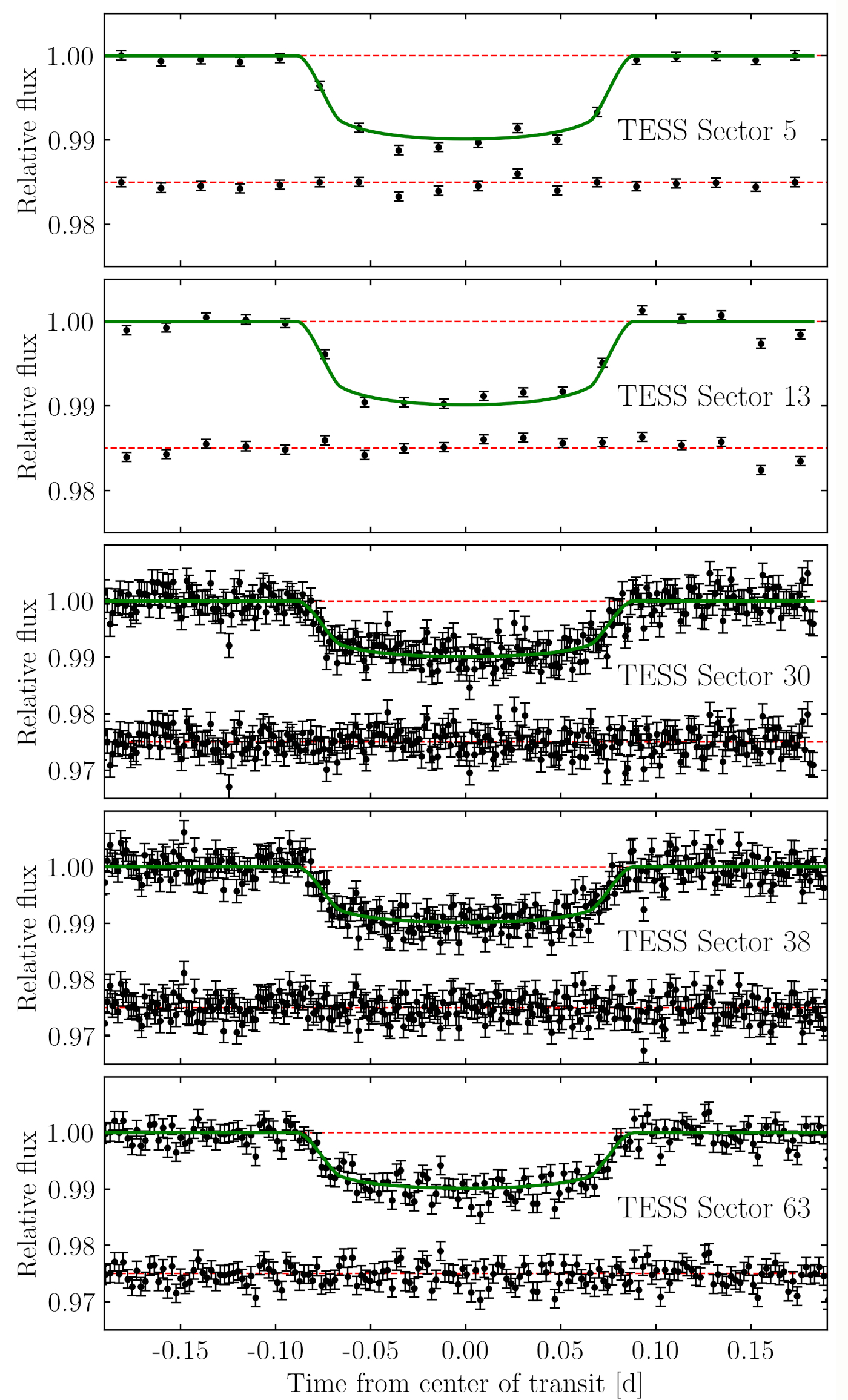}
 \caption{TESS light curves  and respective simultaneous multi-transit
   fit for  five transits  of TOI-4562\,b.   The panels  correspond to
   cycles \#0, \#1, \#3 and \#4 and \#7, from top to bottom. Cycle \#7
   is the as-yet-unpublished  event occurred on March  29, 2023. Phase
   0.00 in the light curves corresponds to the instants of midtransit
   listed in the 1st column of Table 1.}
 \label{fig:all-transits}
\end{figure}

\begin{figure*}
\centering
\includegraphics[scale=0.41]{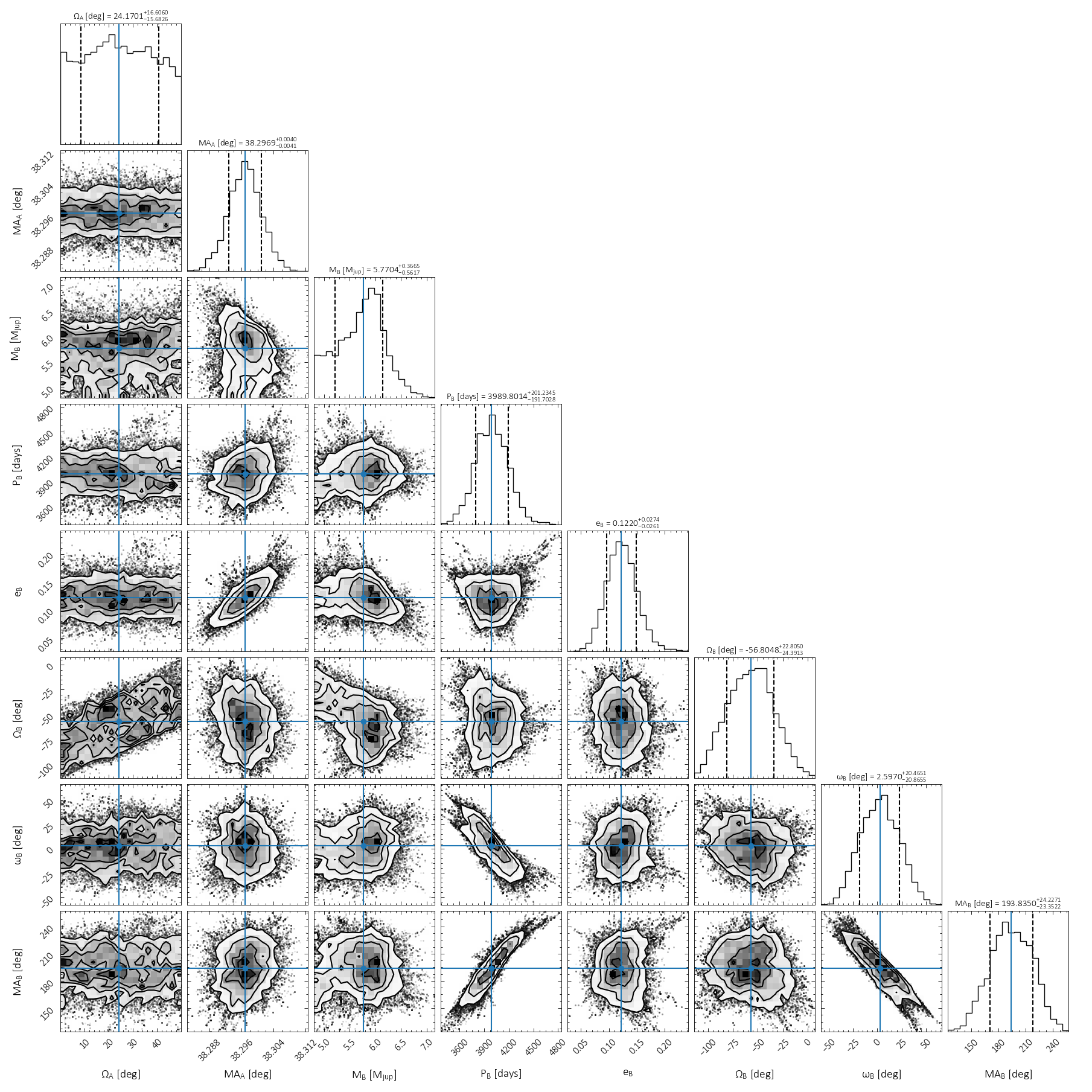}
 \caption{Corner plot  for the  a posterior  distribution of  our free
   orbital parameters  of TOI-4562 b  and TOI-4562 c derived  from our
   analysis. The values for the orbital parameters are listed in Table 2.}
 \label{fig:corner}
\end{figure*}

\begin{figure*}
\sidecaption
\centering
\includegraphics[scale=0.9]{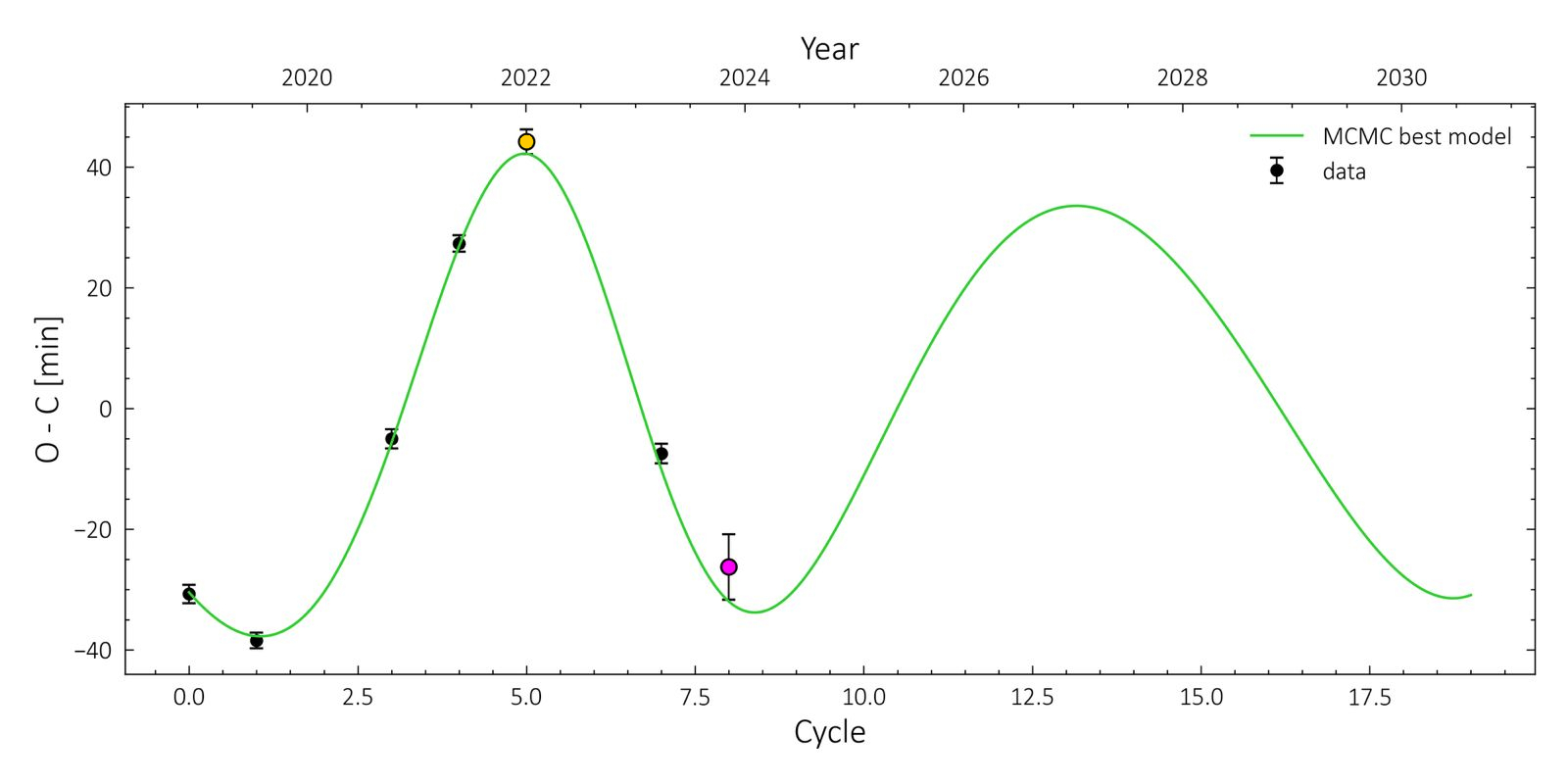}
 \caption{$O-C$ diagram  of the  observed (circles with  errobars) and
   modelled  (solid green  line)  data for  the  already observed  and
   future cycles.  The calculated midtransits  for cycles \#10 to \#15
   are presented in Table A.2.}
 \label{fig:predict}
\end{figure*}

\subsection{Simultaneous multi-transit fit of the TESS observations}

As described  in Section 2  and 3,  we used a  total of 7  transits of
TOI-4562\,b  in  our analysis.   With  the  exception of  the  transit
obtained  at the  LCOGT and  described by  HEI23, for  all other  TESS
transits  we  calculated  the   transit  midpoints  independently,  by
performing        a       simultaneous        multi-transit       fit.
Fig.~\ref{fig:all-transits}    complements    the   fit    shown    in
Fig.~\ref{fig:opd} for  the OPD  transit, showing the  fits for  the 5
TESS  transits, including  the  as-yet-unpublished  event occurred  on
March 29, 2023 (Cycle \#7).

\subsection{Posterior distribution of free parameters}

In Section 4 we describe the  process to obtain the orbital parameters
of TOI-4562\,c, by modelling the posterior probability using the Markov
Chain Monte  Carlo (MCMC) method.  Fig.~\ref{fig:corner}  presents the
posterior  distribution of  our free  parameters: the  third-body mass
($M_{c}$), the  orbital period ($P_{c}$), the  eccentricity ($e_{c}$),
the  longitude of  the ascending  node ($\Omega_{c}$),  the periastron
argument  ($\omega_{c}$),  the  mean anomaly  ($\mathcal{M}{c}$),  the
transiting  body longitude  of the  ascending node  ($\Omega_{b}$) and
mean anomaly ($\mathcal{M}_{b}$).

\subsection{Next events for TOI-4562\,b}

Due to the  uniqueness of TOI-4562 system, transit  times for upcoming
events  of TOI-4562\,b  are presented  in HEI23.   This importance  is
greater  now,  with   the  discovery  of  another   exoplanet  in  the
system. The ephemeris  and TTVs calculated in  this work significantly
changes the instants of the next  transits, which is why the predicted
timing  for  the next  5  transits  of  TOI-4562\,b are  presented  in
Table~\ref{tab:obs}  and   Fig.~\ref{fig:predict}.  As   mentioned  in
Section 4, we  could not find any  sign related to the  transit of the
TOI-4562\,c in the TESS data, but given its long period of more than a
decade and the large orbit, this detection is unlikely.

\begin{table*}
\vspace{36 pt}
\caption{Midtransits for the next TOI-4562\,b events}
\begin{center}
\label{tab:obs}
\begin{tabular}{cccc}
\hline \hline 
\noalign{\smallskip}
Cycle & Transit Mid-time & Transit Date & Visibility \\
      & BJD [days]       &    [UT]          &             \\
\noalign{\smallskip}
\hline
\noalign{\smallskip}
10 & 2,460,707.95681 & 2025-02-01 UT 10:57:47 & Oceania, Antarctica \\
11 & 2,460,933.07844 & 2025-09-14 UT 13:52:57 & Oceania         \\
12 & 2,461,158.19603 & 2026-04-27 UT 16:42:17 & Oceania, Africa \\
13 & 2,461,383.30698 & 2026-12-08 UT 19:22:03 & Africa, Antarctica \\
14 & 2,461,608.41116 & 2027-07-21 UT 21:52:04 & Africa \\
15 & 2,461,833.50982 & 2028-03-03 UT 00:14:08 & South America, Africa \\
\noalign{\smallskip}
\hline
\end{tabular}
\tablefoot{Midtransits  for the  next  five  transits of  TOI-4562\,b
  according   with   the  ephemeris   and   TTVs   described  in   our
  work. Locations in  the Southern Hemisphere where  the transit would
  be observable from ground-based  telescopes are also presented. Full
  or partial coverage will depend on the observatory coordinates.}
\end{center}
\end{table*}

\end{appendix}

\end{document}